\documentclass[twocolumn,twoside]{revtex4}
\usepackage{graphicx}
\usepackage{fancyhdr}
\begin{document}

\title{Experimental study for leptonic and semileptonic decays in the charm sector}

%

\author{S. F. Zhang\\
on behalf of the BESIII Collaboration}
\affiliation{Nanjing University, Jiangsu, China}
\affiliation{Institute of High Energy Physics, Beijing, China}

\begin{abstract}
    Leptonic and semileptonic decays in the charm sector have been well studied
    in recent years. With the largest data sample near $D\bar D$ threshold,
    precision measurements of leptonic and semileptonic decays of charm meson
    and baryon are perfromed at BESIII. Test for letpon flavor universality is
    also performed. Sensitivity for rare leptonic and semileptonic charm decays
    is significantly improved taking advantage of the huge statistics in LHCb
    and the $B$ factories.
\end{abstract}

\maketitle

\thispagestyle{fancy}


\section{Introduction}
Leponic and semileptonic decays are well described in the Standard Model (SM).
The decay amplitudes are proportional to the product of the hadronic
current and the leptonic current. The hadronic current, which describes the
strong interaction in the bound state for leptonic decays and in the hadronic
transition in the semileptonic decays, can not be given directly. Instead, it
can be written in terms of decay constants and form factors. Usually, we have~\cite{Wirbel1985}
\begin{equation}
    H^\mu = <0|-A^\mu|D(p)> = f_Dp^\mu 
\end{equation}
for leptonic decays,
\begin{equation}
    \begin{array}{rcl}
        H^\mu & = & <P(p_2)|V^\mu|D(p_1)> \\
              & = & f_+(q^2)[p_2^\mu-\frac{M_1^2-M_2^2}{q^2}q^\mu]+f_0(q^2)\frac{M_1^2-M_2^2}{q^2}q^\mu 
    \end{array}
\end{equation}
for semileptonic decays to pseudoscalar mesons, and
\begin{equation}
    \begin{array}{rcl}
    H^\mu & = & <V(p_2,\epsilon_2)|V^\mu-A^\mu|D(p_1)> \\
          & = & -(M_1+M_2)\epsilon_2^{*\mu}A_1(q^2)+\frac{\epsilon_2^*q}{M_1+M_2}P^\mu A_2(q^2) \\
          &   & +2M_2\frac{\epsilon_2^*q}{q^2}q^\mu[A_3(q^2)-A_0(q^2)] \\
          &   & +\frac{2i\epsilon_{\mu\nu\rho\sigma}\epsilon^{*\nu}p_1^\rho p_2^\sigma}{M_1+M_2}V(q^2).  
    \end{array}
\end{equation}
for semileptonic decays to vector mesons. Here, $q$ is the total four 
momentum of the lepton system. The charm mesons and baryons are not light
enough for applying perturbation method and not heavy enough for heavy quark
theory, so it's important to measure the decay constants and form factors in experiment.
Since $q^\mu L_\mu$ vanishes when the mass of the lepton equals zero, the contributions from $f_0(q^2)$ and $A_0(q^2)$ are negligible for decays to electron or muon and are usually ignored in experiment.

Meanwhile, some recent results~\cite{Lees2012,Lees2013,Aaij2015,Huschle2015,Sato2016,Aaij2014,Aaij2017,Abdesselam2019} in the measurements of $B$ semileptonic decays
have shown hints of violation of lepton flavor universality (LFU), which requires that
different generations of leptons have the same coupling strength with gauge bosons.
For $D$ meson leptonic decays and semileptonic decays to pseudoscalar mesons, the
hadronic currents are almost cancelled out when we calculate the ratio of decay rates
to different generations of leptons. With these precisely determined SM prediction,
we can test for LFU in the charm sector at a very good precision, which is important
to understand the potential mechanism in the $B$ anomaly.

One can also test the SM in rare charm leptonic and semileptonic decays where the branching
fractions (BFs) may be enhanced by new physics beyond the SM. In particular, the loop diagram
of flavor changing neutral current in the SM is heavily suppressed and may receive contributions
from various kinds of SM extensions.

\section{Facility working on charm leptonic and semileptonic decays}
Measurements of charm leptonic and semileptonic decays are mainly carried out
in the charm factory, nowadays mostly at the BESIII detector~\cite{besiii} with 2.93 fb$^{-1}$
and 3.19 fb$^{-1}$ data collected at center mass energy 3.773 GeV and 4.178 GeV,
respectively. For charm factories
which work at the energy point of $D\bar D$ threshold, the $D$ mesons are produced
in pair. If we first fully reconstruct a $D$ meson using hadronic decays, the other
$D$ meson is then guaranteed to exist in the recoiling system. One can also study
these decays at the $B$ factories, e.~g., Belle~\cite{belle} and BaBar~\cite{babar}, 
where the luminosity is much higher than the charm factories. In this case, the $D$ meson is
often reconstructed from $D^*$ decay, and the flavor of the $D$ meson can be determined
from the accompanying soft pion. Belle has collected about 1 ab$^{-1}$ data near $\Upsilon(4S)$
resonance, and ultimately 50 ab$^{-1}$ are expected at its upgrade, BelleII. BaBar also
collected about 550 fb$^{-1}$ data at the similar energy point.
The LHCb detector~\cite{lhcb1,lhcb2}, working at the Large Hadron Collider, is also contributing data
to charm physics. The statistic is much higher benefiting from the huge cross section
at the hadron collider. However, the electron is hard to reconstruct due to bremsstrahlung
and it is very difficult to reconstruct decay channels involving neutrinos.
In 2018, LHCb has finished its run 2 data taking and
in total about 9 fb$^{-1}$ data has been collected.
Compared to the charm factories, study of charm leptonic and semileptonic decays at
the $B$ factories and LHCb suffers from lower efficiency and higher backgrounds. However,
the high statistics make them superior places for rare decay search.

\section{Recent results}

\subsection{Leptonic decays}
\begin{figure}[htbp]\centering
    \includegraphics[width=0.45\textwidth]{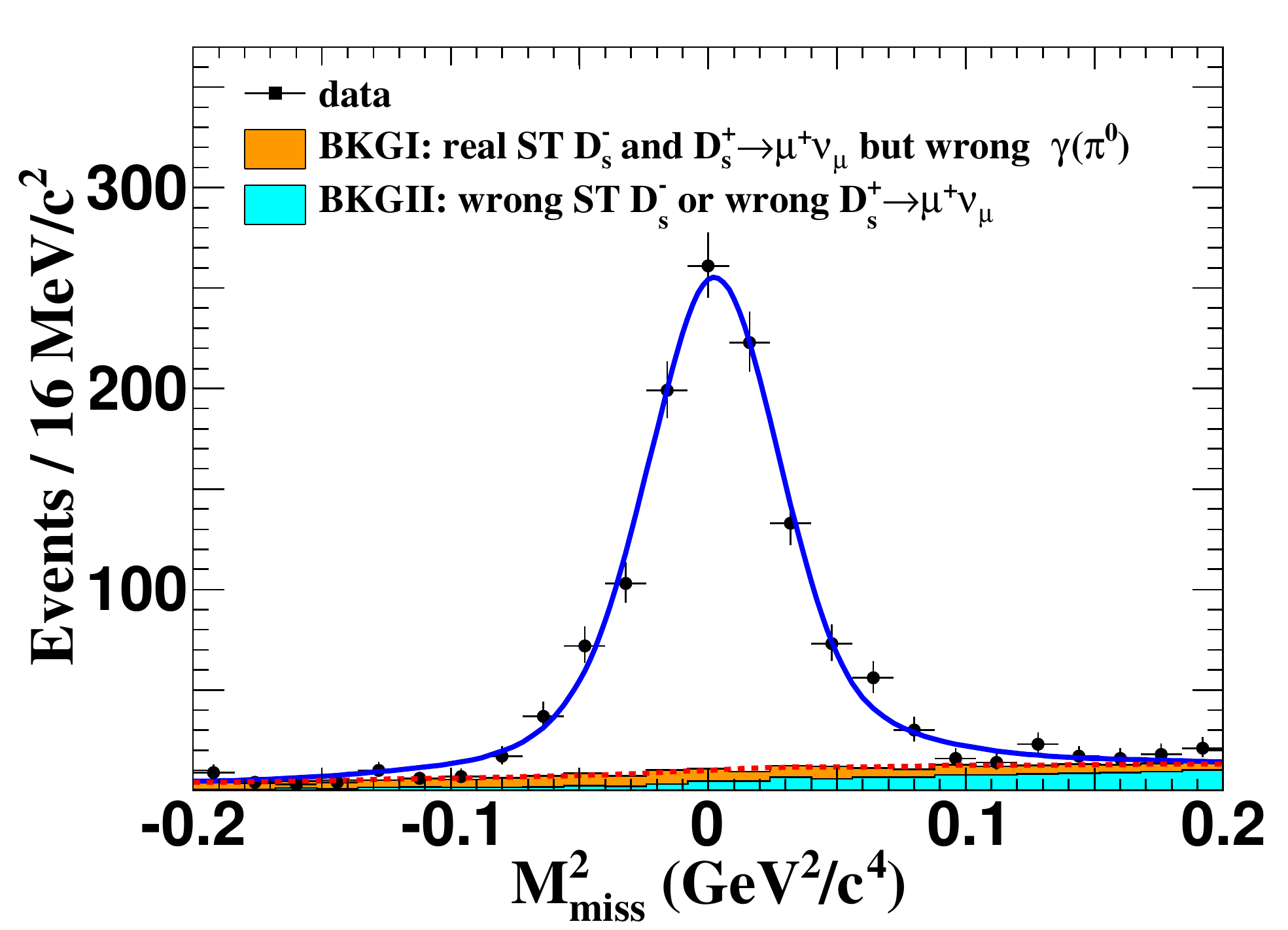}
    \caption{\label{fig:Dstomuv}Fit to the $M_{\rm miss}^2$ distribution of
    $D_s^+\to\mu^+\nu_\mu$ candidates, where the dots with error bars are
    data, the blue solid curve shows the best fit and the red dashed curve
    shows the background shape.}
\end{figure}

Leptonic decays give us direct access to the corresponding decay constants and 
CKM matrix elements. BESIII has recently published the measurement of the BF
of $D_s^+\to\mu^+\nu_\mu$~\cite{bes_dstomuv}. Figure~\ref{fig:Dstomuv} shows the missing mass
square of the neutrino, which is defined as $M_{\rm miss}^2=\sqrt{E_{\rm miss}^2-
p_{\rm miss}^2}$, where $E_{\rm miss}$ and $p_{\rm miss}$ are the missing energy and
momentum of the candidate event. The BF is measured to be
$$\mathcal{B}(D_s^+\to\mu^+\nu_\mu)=(5.49\pm0.16\pm0.15)\times10^{-3}.$$
Given the lifetime of $D_s^+$ meson~\cite{PDG2018}, one obtains
$$f_{D_s^+}|V_{cs}|=246.2\pm3.6\pm3.5~{\rm MeV}.$$
If we input the CKM matrix element $|V_{cs}|$ from a global fit~\cite{PDG2018}, we have
$$f_{D_s^+}=252.9\pm3.7\pm3.5~{\rm MeV}.$$
In contrast, inputting $f_{D_s^+}$ from Lattice QCD calculation~\cite{Bazavov2018,Carrasco2015} gives
$$|V_{cs}|=0.985\pm0.014\pm0.014.$$
In any case, these are the most precise single measurements to date.
Averaging this measurement of $\mathcal{B}(D_s^+\to\mu^+\nu_\mu)$ with the previous
measurements~\cite{cleo_dstomuv,babar_dstomuv,belle_dstomuv,bes_dstolv} and combining the world average of $\mathcal{B}(D_s^+\to\tau^+\nu_\tau)$~\cite{PDG2018},
we find
$$\frac{\mathcal{B}_{D_s^+\to\tau^+\nu_\tau}}{\bar {\mathcal{B}}_{D_s^+\to\mu^+\nu_\mu}}=9.98\pm0.52,$$
which is consistent with the SM prediction 9.74.

\begin{figure}[htbp]\centering
    \includegraphics[width=0.45\textwidth]{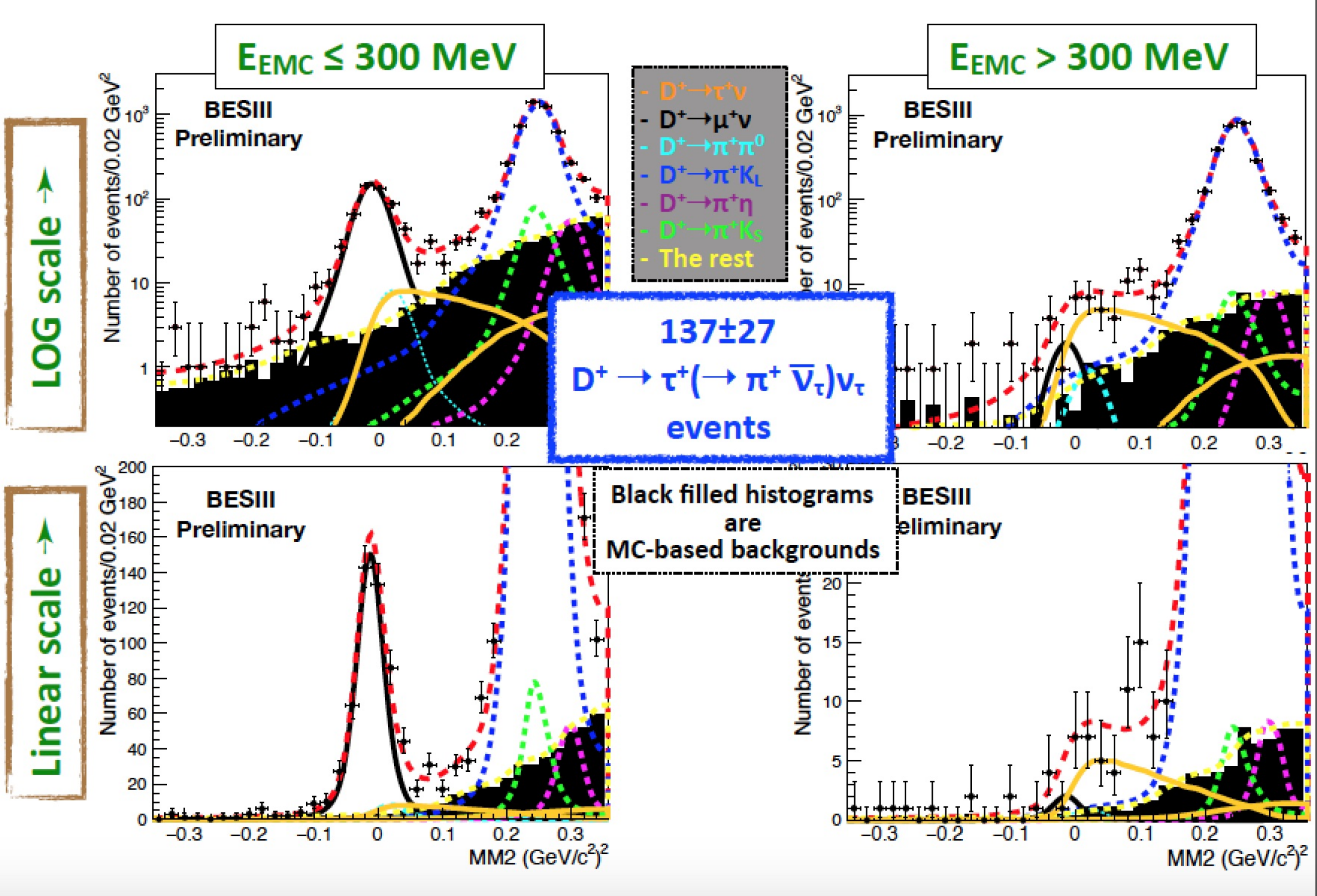}
    \caption{\label{fig:dptotauv}Fit to $M_{\rm miss}^2$ of $D^+\to\tau^+\nu_\tau$ candidates.}
\end{figure}

BESIII has also released the preliminary result of
the search for $D^+\to\tau^+\nu_\tau$, as shown in
Fig.~\ref{fig:dptotauv}. The data sample is divided
according to the energy deposited in the electromagnetic calorimeter into pion and muon dominated samples to better constrain the background from $D^+\to\mu^+\nu_\mu$. We have
$$\mathcal{B}(D^+\to\tau^+\nu_\tau)=(1.20\pm0.24)\times10^{-3},$$
and 
$$f_{D^+}|V_{cd}|=50.4\pm5.0~{\rm MeV}$$
using the world average $D^+$ lifetime~\cite{PDG2018}. 
Here the uncertainties are statistical only and
the statistical significance is 4$\sigma$.

Combined with the previous measurement of the BF
of $D^+\to\mu^+\nu_\mu$ from BESIII~\cite{bes_dptomuv}, we
obtain
$$\frac{\mathcal{B}_{D^+\to\tau^+\nu_\tau}}{{\mathcal{B}}_{D^+\to\mu^+\nu_\mu}}=3.21\pm0.64,$$
which is also consistent with the SM prediction 2.66.

\subsection{Semileptonic decays}
$D$ meson semileptonic decays to pseudoscalar mesons can be fully characterized by the single kinematic variable $q^2$. 
The form factors, which describe the hadronic transition, are usually parametrized as the function of $q^2$ with respect to its value $q^2=0$. 
The most widely used parametrizations include the Single Pole Model~\cite{Wirbel1985}, the Modified Pole Model~\cite{mpm}, the ISGW2 Model~\cite{isgw2} and the series expansion parametrization~\cite{sep}.
In experiment, one usually divide the $q^2$ distribution into several intervals and measure the partial decay rate in each interval. The measured partial decay rates are then fitted using the theoretically expected form to extract the form factor at $q^2=0$.

\begin{figure}[htbp]\centering
    \includegraphics[width=0.45\textwidth]{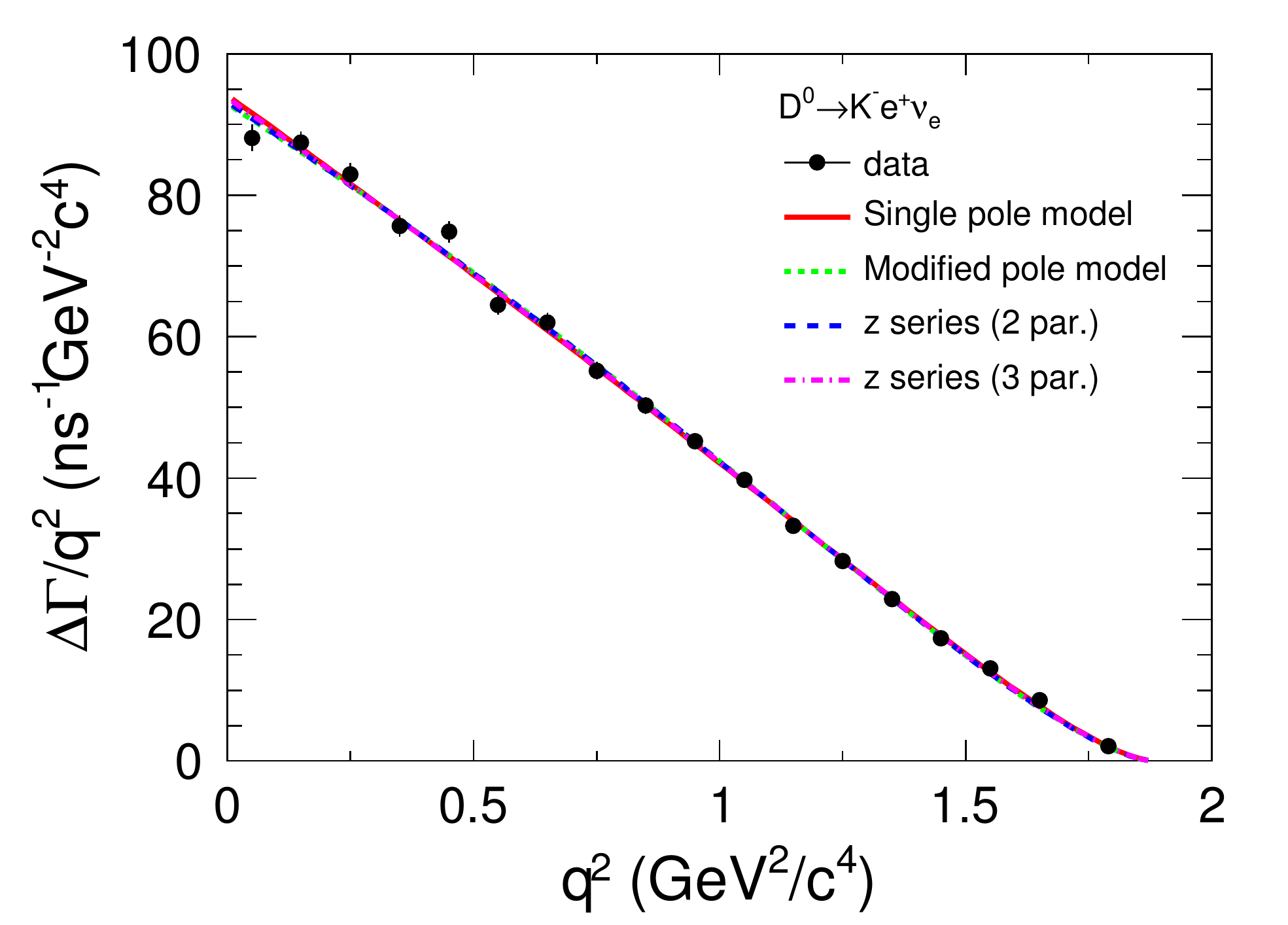}
    \caption{\label{fig:d0tokev}Fit to partial decays rates in 18 $q^2$ intervals for $D^0\to K^-e^+\nu_e$.}
\end{figure}

One such example is shown in Fig.~\ref{fig:d0tokev} for $D^0\to K^-e^+\nu_e$ published by BESIII~\cite{bes_d0tok_pi_ev}. 
In recent years, similar analyses have been performed
at BESIII for $D\to\pi e^+\nu_e$, $D^0\to K^-\mu^+\nu_\mu$, and $D\to\eta(^\prime)e^+\nu_e$, etc~\cite{bes_d0tok_pi_ev,bes_d0tokmuv,bes_dptok_pi_ev,bes_dptoetaev,bes_dstoetaev,bes_dstokev}. The measured BFs and the product of form factors at $q^2=0$ and CKM matrix elements are summarized in Table~\ref{tab:psuedoscalar}. Here the results listed in the table are extracted using the 2-parameter series expansion parametrization.
\begin{table*}[htbp]\centering
    \caption{\label{tab:psuedoscalar}The measured BFs and product of form factors at $q^2=0$ and CKM matrix elements for some $D$ meson semileptonic decays to psuedoscalar mesons.}
    \begin{tabular}{lc|lc}\hline
        $\mathcal{B}(D^0\to K^-e^+\nu_e)$ & $(3.505\pm0.014\pm0.033)\%$ & $f_+^{D^0\to K^-}(0)|V_{cs}|$ & $0.7172\pm0.0025\pm0.0035$ \\
        $\mathcal{B}(D^0\to K^-\mu^+\nu_\mu)$ & $(3.431\pm0.019\pm0.035)\%$ & $f_+^{D^0\to K^-}(0)|V_{cs}|$ & $0.7133\pm0.0038\pm0.0030$ \\
        $\mathcal{B}(D^+\to\bar K^0e^+\nu_e)$ & $(8.60\pm0.06\pm0.015)\%$ & $f_+^{D^+\to\bar K^0}(0)|V_{cs}|$ & $0.7053\pm0.0040\pm0.0112$ \\
        $\mathcal{B}(D_s^+\to K^0e^+\nu_e)$ & $(3.25\pm0.38\pm0.16)\times10^{-3}$ & $f_+^{D_s^+\to K^0}(0)|V_{cd}|$ & $0.162\pm0.019\pm0.003$ \\
        $\mathcal{B}(D^0\to\pi^-e^+\nu_e)$ & $(2.95\pm0.04\pm0.03)\times10^{-3}$ & $f_+^{D^0\to\pi^-}(0)|V_{cd}|$ & $0.1435\pm0.0018\pm0.0009$ \\
        $\mathcal{B}(D^+\to\pi^0e^+\nu_e)$ & $(3.63\pm0.08\pm0.05)\times10^{-3}$ & $f_+^{D^+\to\pi^0}(0)|V_{cd}|$ & $0.1400\pm0.0026\pm0.0007$ \\
        $\mathcal{B}(D^+\to\eta e^+\nu_e)$ & $(10.74\pm0.81\pm0.51)\times10^{-4}$ & $f_+^{D^+\to\eta}(0)|V_{cd}|$ & $0.0786\pm0.0064\pm0.0021$ \\
        $\mathcal{B}(D_s^+\to\eta e^+\nu_e)$ & $(2.323\pm0.063\pm0.063)\%$ & $f_+^{D_s^+\to\eta}(0)|V_{cs}|$ & $0.4455\pm0.0053\pm0.0044$ \\
        $\mathcal{B}(D_s^+\to\eta^\prime e^+\nu_e)$ & $(0.824\pm0.073\pm0.027)\%$ & $f_+^{D_s^+\to\eta^\prime}(0)|V_{cs}|$ & $0.477\pm0.049\pm0.011$ \\
        \hline
    \end{tabular}
\end{table*}
Most of the measurements are consistent with the LQCD calculations~\cite{Na2010,Na2011,Lubicz2017,Li2019}.
We can also find that
$$\frac{f_+^{D_s^+\to K^0}(0)}{f_+^{D^+\to\pi^0}(0)}=1.16\pm0.14\pm0.02,$$
which is consistent with U-spin symmetry.

\begin{figure*}[htbp]\centering
    \includegraphics[width=0.45\textwidth]{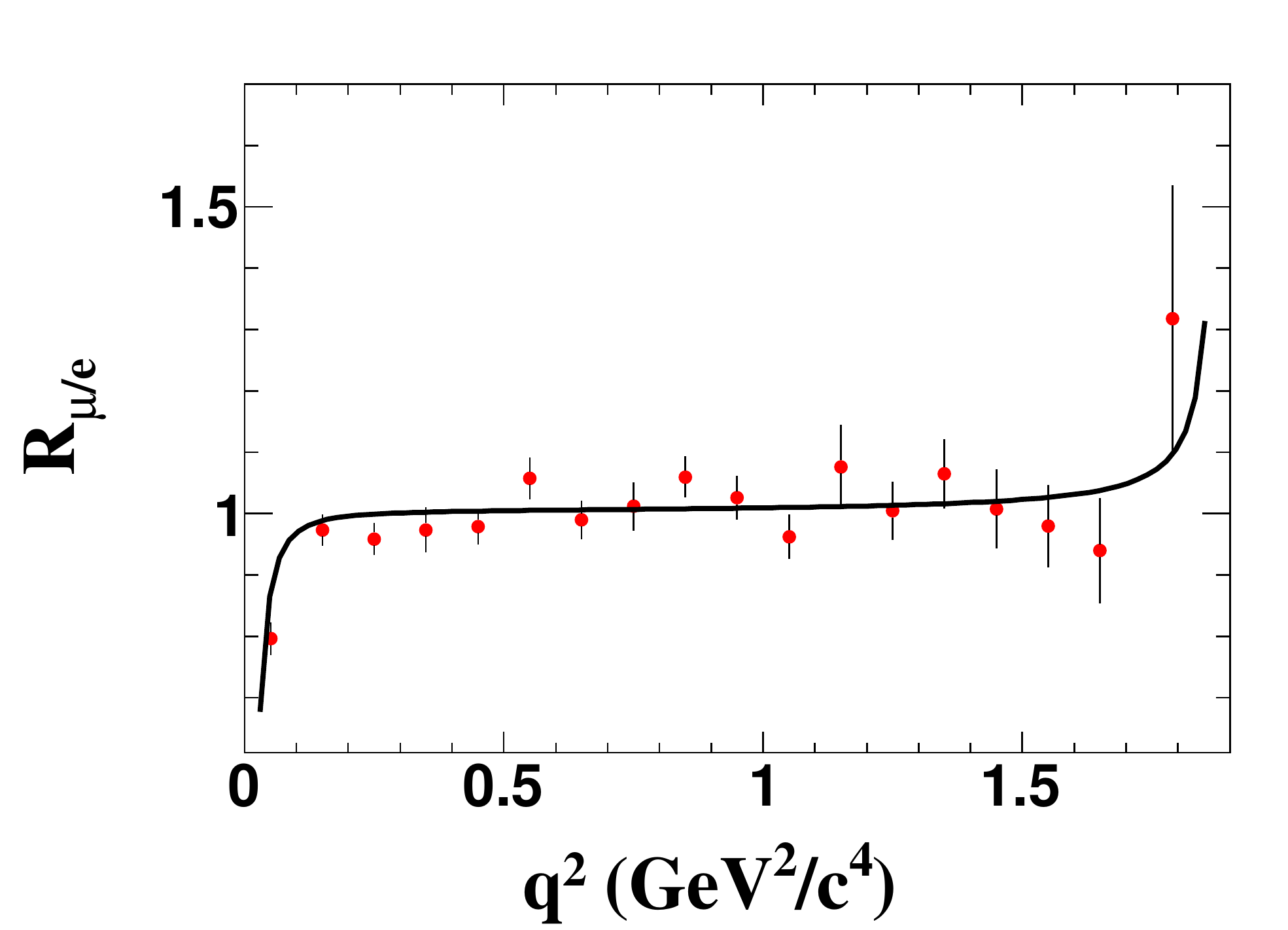}
    \includegraphics[width=0.45\textwidth]{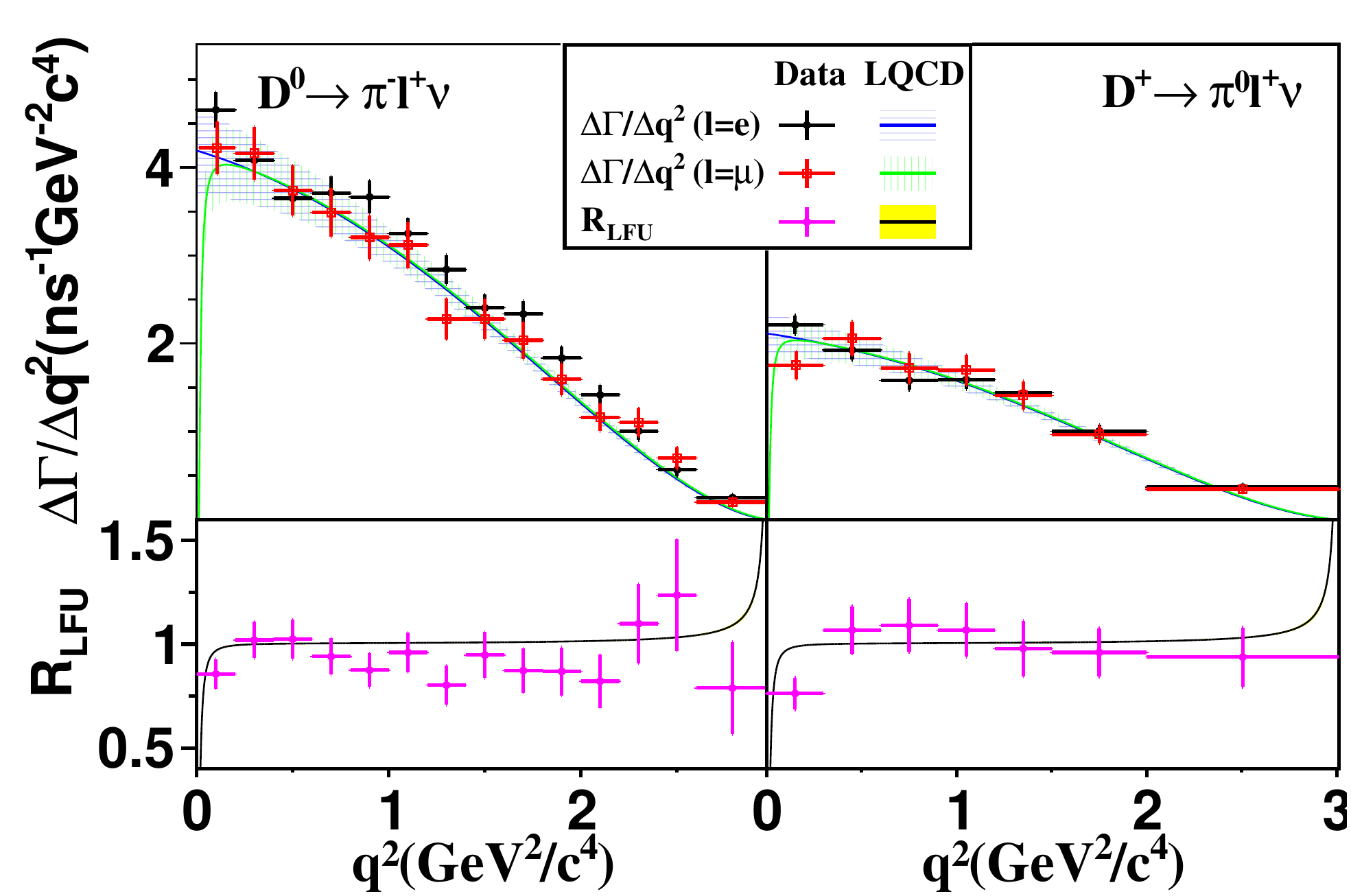}
    \put(-350,120){$D^0\to K^-\ell^+\nu_\ell$}
    \caption{\label{fig:lfu}Test for LFU in different $q^2$ intervals, where the points with error bars are data and the solid curves are the SM predictions.}
\end{figure*}

As introduced before, the ratio of the decay rates for $D$ meson semileptonic decays to pseudoscalar mesons with muon or with electron can be precisely determined in SM theory, which makes it an ideal place for test of LFU.
Such studies have been performed at BESIII, yielding~\cite{bes_d0tokmuv,bes_dptokmuv,bes_dtopimuv}
$$\frac{\Gamma(D^0\to K^-\mu^+\nu_\mu)}{\Gamma(D^0\to K^-e^+\nu_e)}=0.974\pm0.014,$$
$$\frac{\Gamma(D^+\to\bar K^0\mu^+\nu_\mu)}{\Gamma(D^+\to\bar K^0e^+\nu_e)}=1.014\pm0.017,$$
$$\frac{\Gamma(D^0\to\pi^-\mu^+\nu_\mu)}{\Gamma(D^0\to\pi^-e^+\nu_e)}=0.922\pm0.037,$$
and 
$$\frac{\Gamma(D^+\to\pi^0\mu^+\nu_\mu)}{\Gamma(D^+\to\pi^0e^+\nu_e)}=0.964\pm0.045,$$
These are all consistent with the SM prediction within 2.5$\sigma$~\cite{Riggio2018}.
One may wish to look at the ratios in different $q^2$ intervals which give us better
sensitivity to potential LFU violation effect. Figure~\ref{fig:lfu} shows the result
for $D^0\to K^-\ell^+\nu_\ell$, $D^0\to\pi^-\ell^+\nu_\ell$, and $D^+\to\pi^0\ell^+\nu_\ell$,
where the theoretical expectations are taken from a recent LQCD calculation from ETM
Collaboration~\cite{Lubicz2017}. No significant deviation is observed.

\begin{figure}[htbp]\centering
    \includegraphics[width=0.45\textwidth]{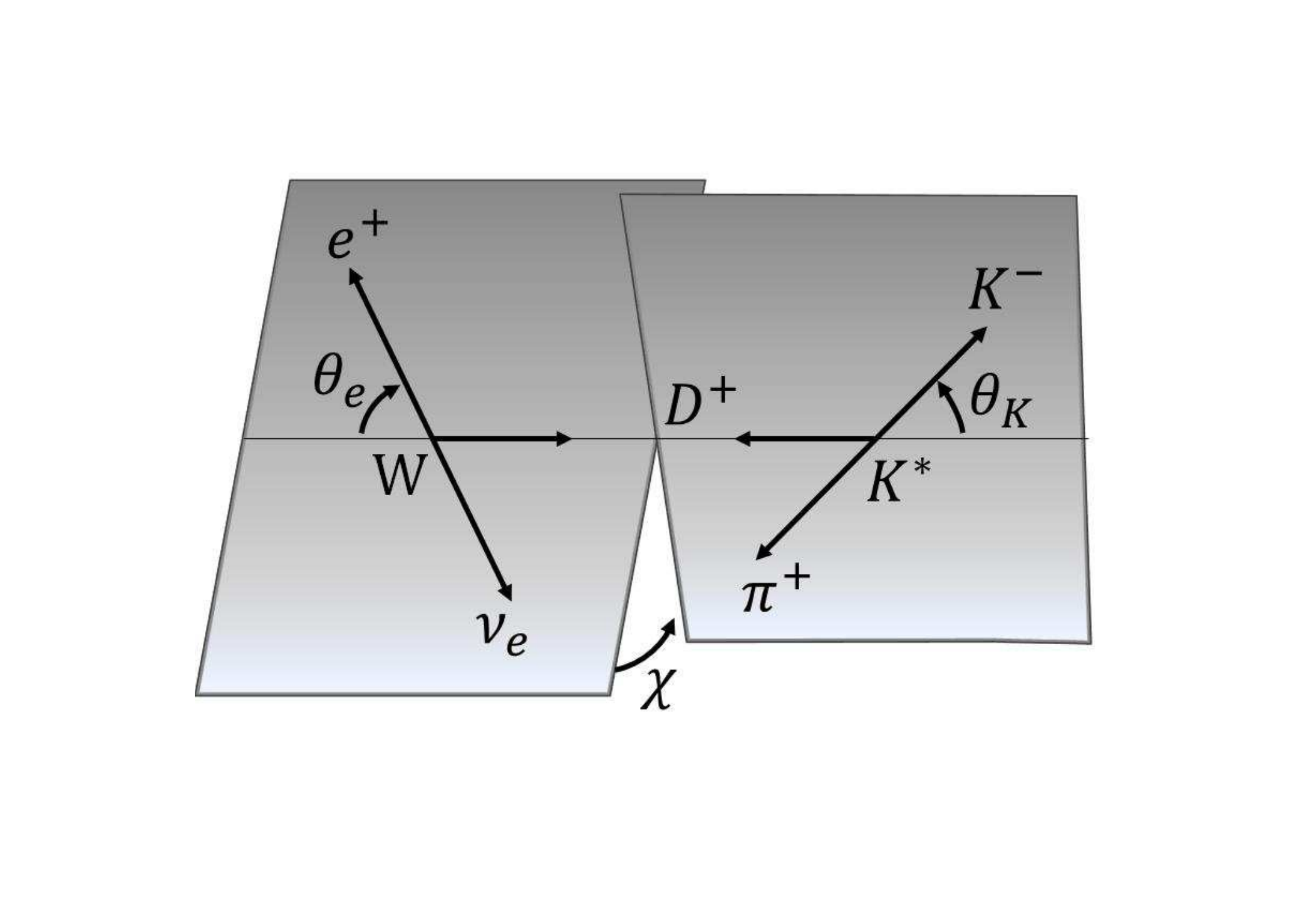}
    \caption{\label{fig:angular}Definition of the angular variables.}
\end{figure}

The situation is more complex in the case of semileptonic decays to vector mesons with the presence of extra polarization vectors.
As illustrated in Fig.~\ref{fig:angular} for $D^+\to\bar K^{*0}e^+\nu_e$~\cite{bes_dptokpiev}, the decay rate is described by three extra angular variables in addition to $q^2$, including the angle between the $\pi$ and the $D$ direction in the $K\pi$ rest frame ($\theta_K$), the angle between the $\nu_e$ and the $D$ direction in the $e\nu_e$ rest frame ($\theta_e$), and the angle between the two decay planes ($\chi$).

\begin{figure}[htbp]\centering
    \includegraphics[width=0.23\textwidth]{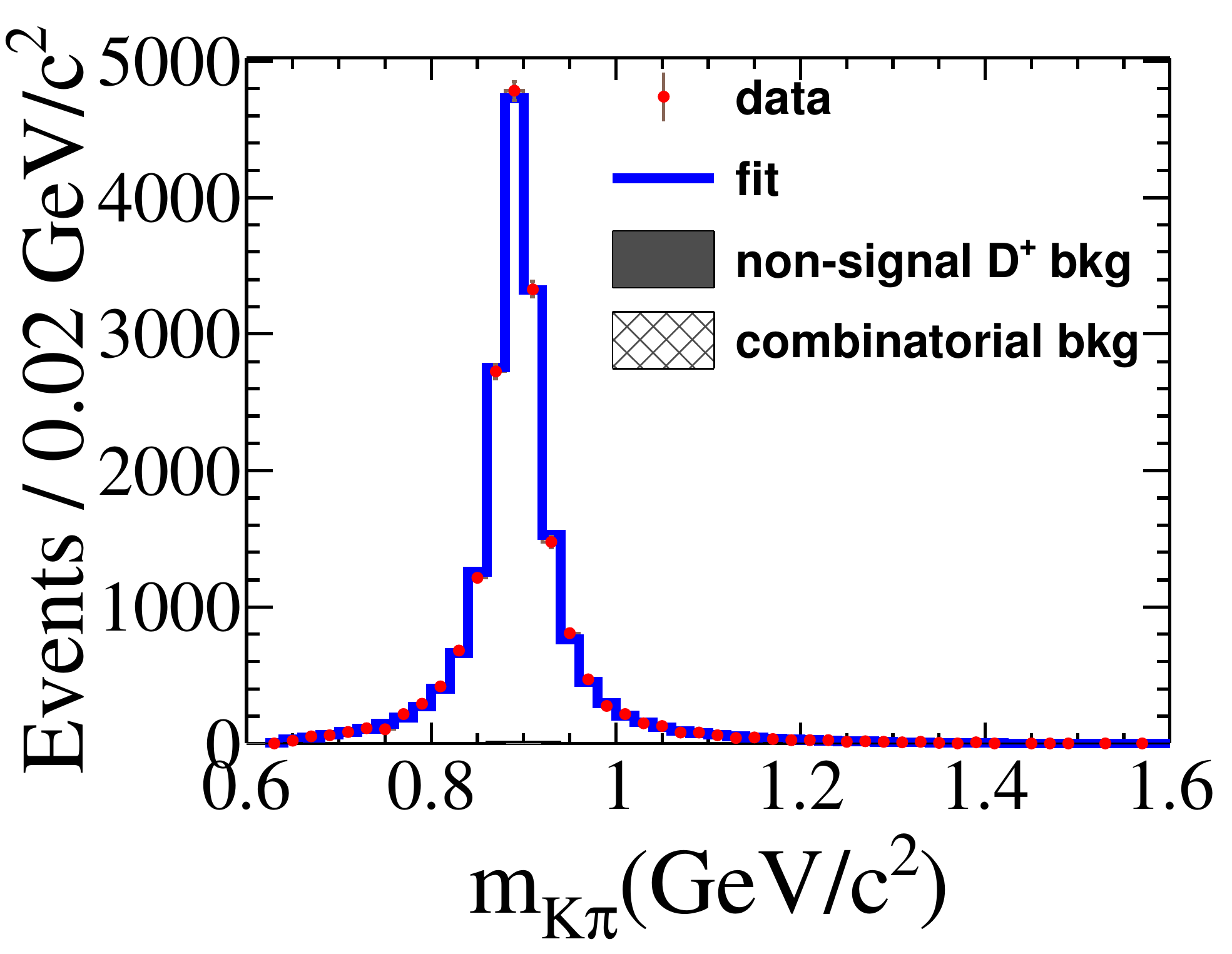}
    \includegraphics[width=0.23\textwidth]{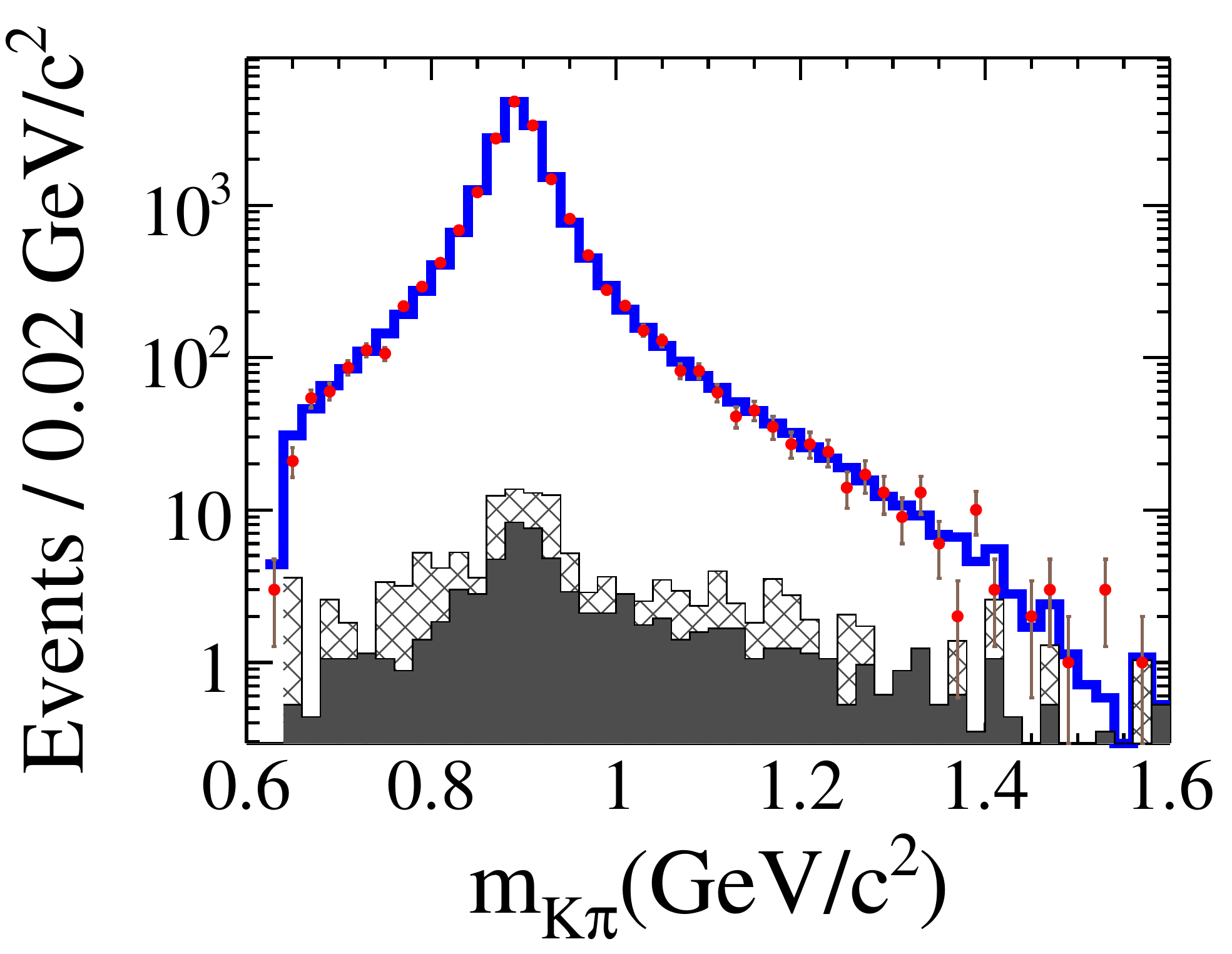}\\
    \includegraphics[width=0.23\textwidth]{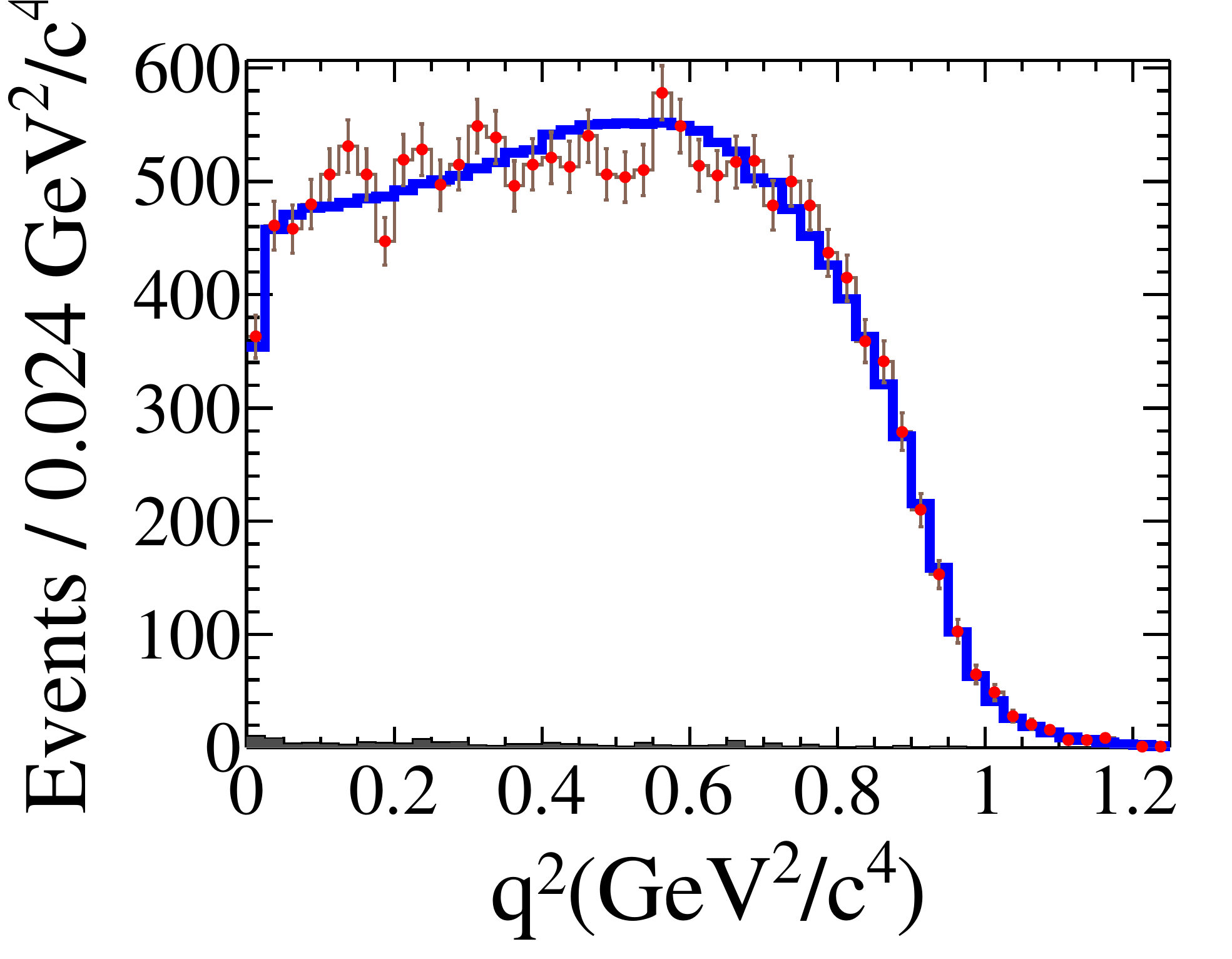}
    \includegraphics[width=0.23\textwidth]{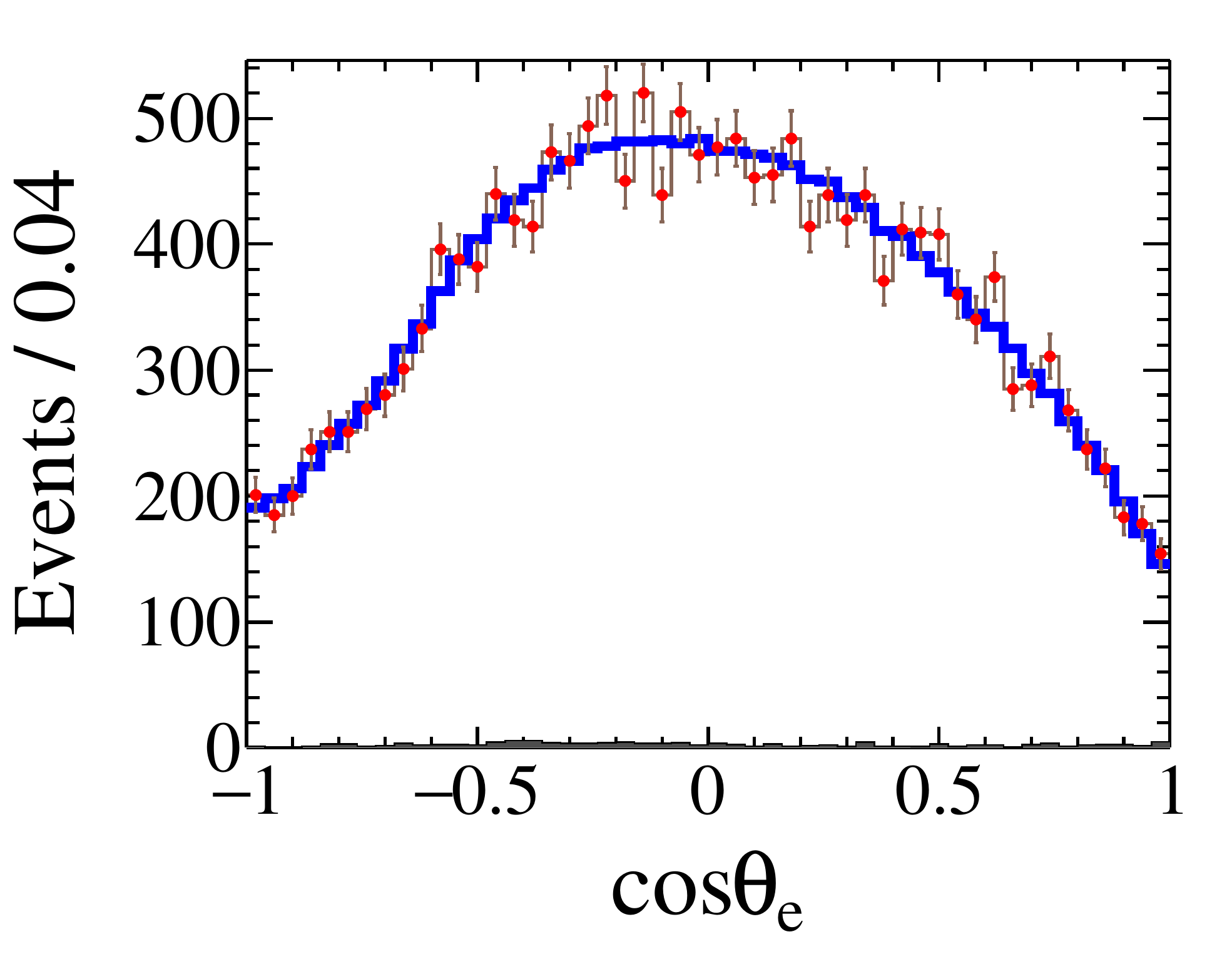}\\
    \includegraphics[width=0.23\textwidth]{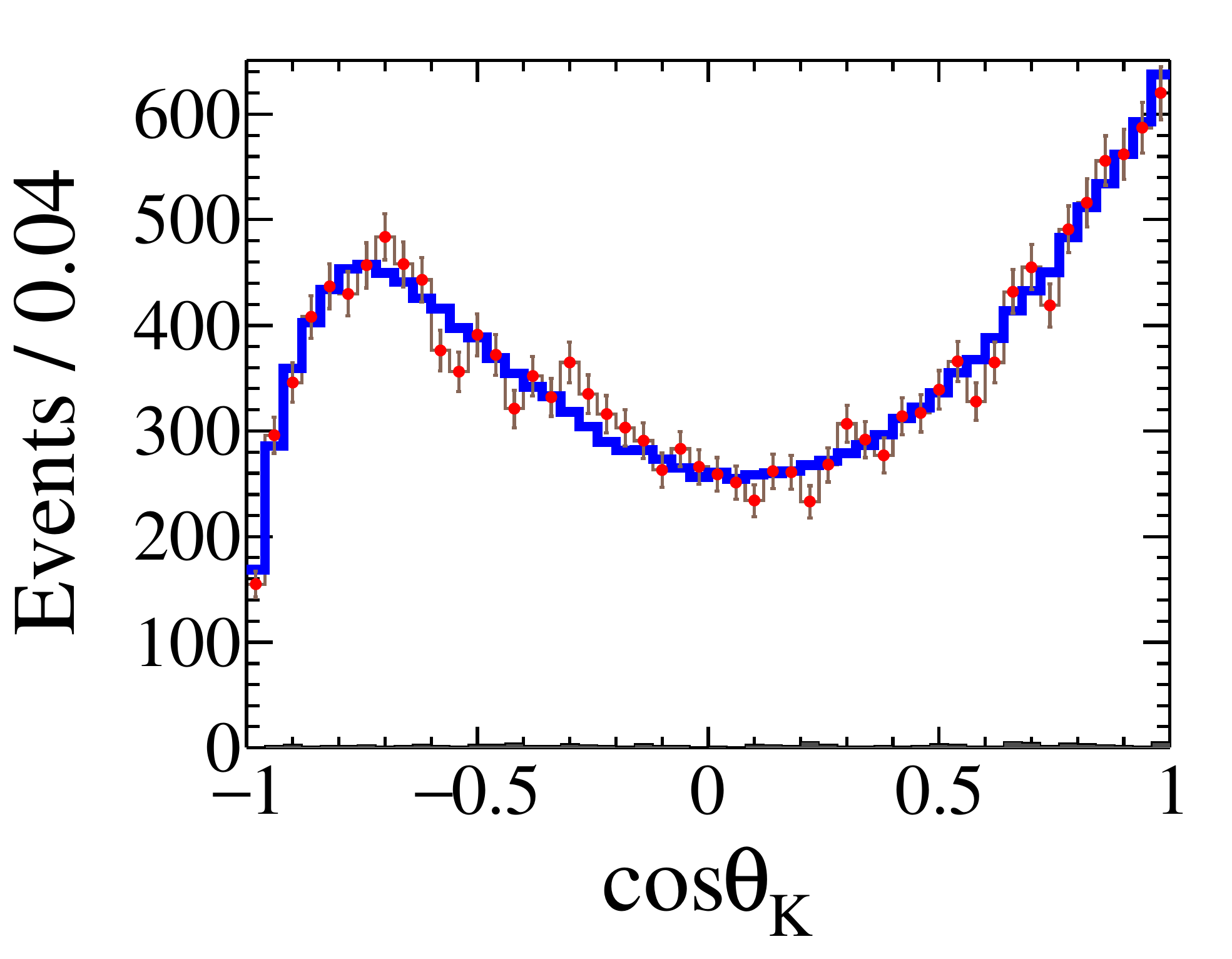}
    \includegraphics[width=0.23\textwidth]{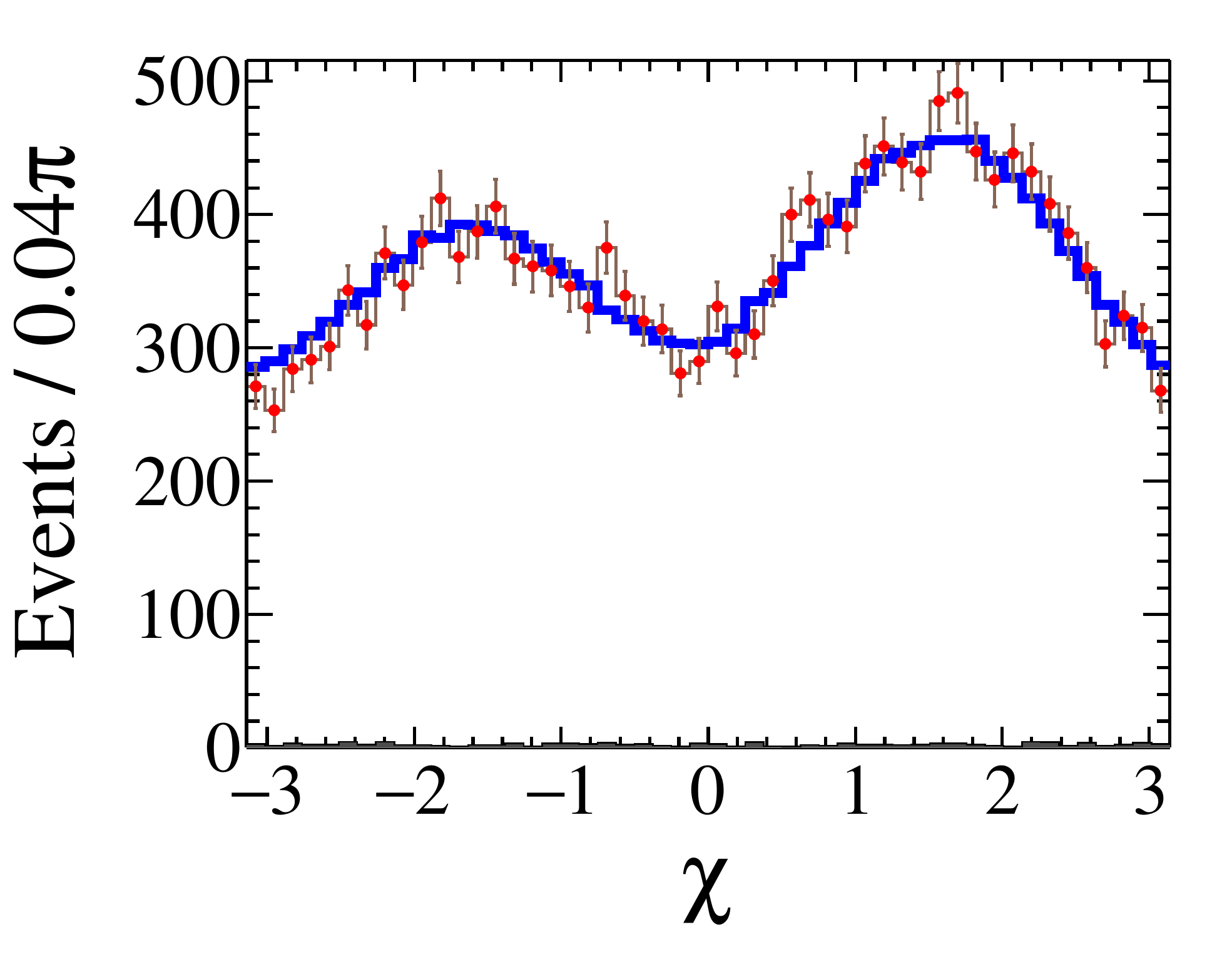}
    \caption{\label{fig:DptoKpiev}Projections of the fit to the five kinematic variables.}
\end{figure}

Meanwhile, we now have four form factors $A_{0,1,2}(q^2)$ and $V(q^2)$ ($A_3(q^2)$ is not independent), where $A_0(q^2)$ is often ignored. Due to the limited statistics, these form factors are usually modeled using the Single Pole Model. What's more, since $A_1(q^2)$ appears in every helicity amplitude, we often measure the form factor ratios
$$r_V=\frac{V(0)}{A_1(0)},$$
and
$$r_2=\frac{A_2(0)}{A_1(0)},$$
as they do not require additional inputs like $D$ meson lifetime and CKM matrix elements.
Figure~\ref{fig:DptoKpiev} shows the fit on $q^2$ and the three angular variable distributions, as well as the $K\pi$ invariant mass distribution for $D^0\to K^-\pi^+e^+\nu_e$ candidate events.
The BF of the $\bar K^{*0}$ contribution and the extracted form factor ratios are summarized in Table~\ref{tab:vector}. With the input of $D^+$ lifetime~\cite{PDG2018} and $|V_{cs}|$~\cite{PDG2018}, we obtain
$$A_1(0)=0.589\pm0.010\pm0.012.$$
Similar analyses are also performed for $D^0\to\bar K^0\pi^-e^+\nu_e$~\cite{bes_d0tokpiev}, $D^+\to\omega e^+\nu_e$~\cite{bes_dptoomegaev}, $D\to\pi\pi e^+\nu_e$~\cite{bes_dtopipiev} and $D_s^+\to K^{*0}e^+\nu_e$~\cite{bes_dstokev} at BESIII and the results are summarized in Table~\ref{tab:vector}.
We also notice that significant S-wave contribution is observed for $f_0(500)$ in $D^+\to\pi^-\pi^+e^+\nu_e$ with
$\mathcal{B}(D^+\to f_0(500)e^+\nu_e, f_0(500)\to\pi^+\pi^-)=(0.630\pm0.043\pm0.032)\times10^{-3}$,
while no evidence for $D^+\to f_0(980)e^+\nu_e$ is observed.
\begin{table*}[htbp]\centering
    \caption{\label{tab:vector}The measured BFs and form factor ratios for some of the $D$ meson decays to vector mesons.}
    \begin{tabular}{lccc}\hline
        Decay & BF & $r_V$ & $r_2$ \\\hline
        $D^+\to\bar K^{*0}e^+\nu_e$ & $(3.54\pm0.03\pm0.08)\%$ & $1.41\pm0.06\pm0.01$ & $0.79\pm0.04\pm0.01$ \\
        $D^0\to\bar K^{*-}e^+\nu_e$ & $(1.36\pm0.03\pm0.03)\%$ & $1.46\pm0.07\pm0.02$ & $0.67\pm0.06\pm0.01$ \\
        $D^+\to\omega e^+\nu_e$ & $(1.63\pm0.11\pm0.08)\times10^{-3}$ & $1.24\pm0.09\pm0.06$ & $1.06\pm0.15\pm0.05$ \\
        $D^0\to\rho^-e^+\nu_e$ & $(1.45\pm0.05\pm0.04)\times10^{-3}$ & $1.70\pm0.08\pm0.05$ & $0.85\pm0.06\pm0.04$ \\
        $D^+\to\rho^0e^+\nu_e$ & $(1.86\pm0.07\pm0.06)\times10^{-3}$ & $1.70\pm0.08\pm0.05$ & $0.85\pm0.06\pm0.04$ \\
        $D_s^+\to K^{*0}e^+\nu_e$ & $(2.37\pm0.26\pm0.20)\times10^{-3}$ & $1.67\pm0.34\pm0.16$ & $0.77\pm0.28\pm0.07$ \\
        \hline
    \end{tabular}
\end{table*}
The U-spin symmetry is also found be conserved with
$$\frac{r_V^{D_s^+\to K^{*0}}}{r_V^{D^+\to\rho^0}}=1.13\pm0.26\pm0.11,$$
$$\frac{r_2^{D_s^+\to K^{*0}}}{r_2^{D^+\to\rho^0}}=0.93\pm0.36\pm0.10.$$

It is also interesting to search for $D$ meson semileptonic decays to scalar mesons which may help us understand the internal structure of the light scalar mesons. 
BESIII has recently searched for $D\to a_0(980)e^+\nu_e$~\cite{bes_dtoa0ev}, where $D^0\to a_0(980)^-e^+\nu_e$ is observed for the first with 6.4$\sigma$ significance. 
The BF is measured to be $\mathcal{B}(D^0\to a_0(980)^-e^+\nu_e, a_0(980)^-\to\eta\pi^-)=(1.33^{+0.33}_{-0.29}\pm0.09)\times10^{-4}$.
The significance for $D^+\to a_0(980)^0e^+\nu_e$ is 2.9$\sigma$ with $\mathcal{B}(D^+\to a_0(980)^0e^+\nu_e, a_0(980)^0\to\eta\pi^0)=(1.66^{+0.81}_{-0.66}\pm0.11)\times10^{-4}$, and less than $3.0\times10^{-4}$ at 90\% confidence level.
Ref.~\cite{Wang2010} proposed a model-independent method to study the nature of light scalar mesons with
$$R=\frac{{\mathcal B}(D^+\to f_0(980)e^+\nu_e)+{\mathcal B}(D^+\to f_0(500)e^+\nu_e)}{{\mathcal B}(D^+\to a_0(980)^0e^+\nu_e)}.$$
$R$ is estimated to be $1.0\pm0.3$ for two-quark description of these scalar mesons, and $3.0\pm0.9$ for tetraquark description.
With BESIII's results~\cite{bes_dtopipiev} we have $R>2.7$ at 90\% confidence level, which favors the SU(3) nonet tetraquark description of the $f_0(500)$, $f_0(980)$, and $a_0(980)$.

The first measurements of the absolute BFs of charm baryon semileptonic decays are also
performed at BESIII using 567 fb$^{-1}$ data taken at center mass energy $\sqrt{s}=4.6$ GeV~\cite{bes_lambdaev,bes_lambdamuv}.
We find 
$$\mathcal{B}(\Lambda_c^+\to\Lambda e^+\nu_e)=(3.63\pm0.38\pm0.20)\%,$$
$$\mathcal{B}(\Lambda_c^+\to\Lambda\mu^+\nu_\mu)=(3.49\pm0.46\pm0.26)\%,$$
and
$$\frac{\Gamma(\Lambda_c^+\to\Lambda\mu^+\nu_\mu)}{\Gamma(\Lambda_c^+\to\Lambda e^+\nu_e)}=0.96\pm0.16\pm0.04.$$
These are consistent with a recent LQCD calculation~\cite{Meinel2016}.

\subsection{Rare decays}
Some decays which are heavily suppressed in the SM may be enhanced by new physics mechanisms, which may be within the sensitivity of current experiments.

BESIII has searched for the radiative leptonic decays $D\to\gamma e^+\nu_e$~\cite{bes_dptogev,bes_dstogev}. Unlike the pure leptonic decays $D\to e^+\nu_e$, these decays do not subject to helicity suppression and the BFs are predicted to be about $10^{-4}$-$10^{-5}$~\cite{Geng2000,Lu2003,Yang2012,Yang2014}.
No signal is observed at BESIII with photon energy larger than 10 MeV and the upper limits are set at 90\% confidence level with
$$\mathcal{B}(D^+\to\gamma e^+\nu_e)<3.0\times10^{-5},$$
$$\mathcal{B}(D_s^+\to\gamma e^+\nu_e)<1.3\times10^{-4}.$$

\begin{figure}[htbp]\centering
    \includegraphics[width=0.23\textwidth]{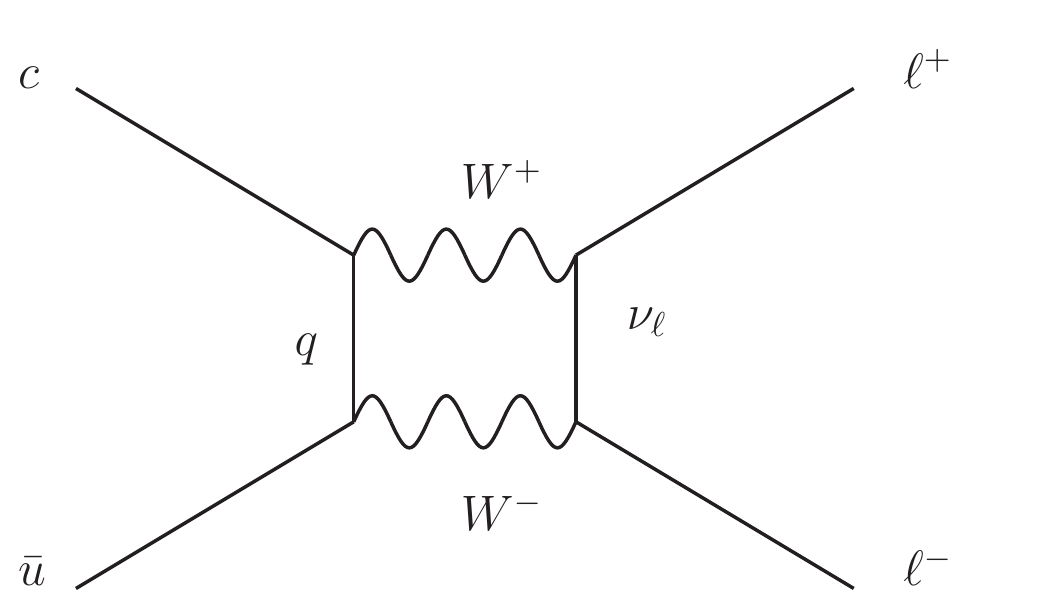}
    \includegraphics[width=0.23\textwidth]{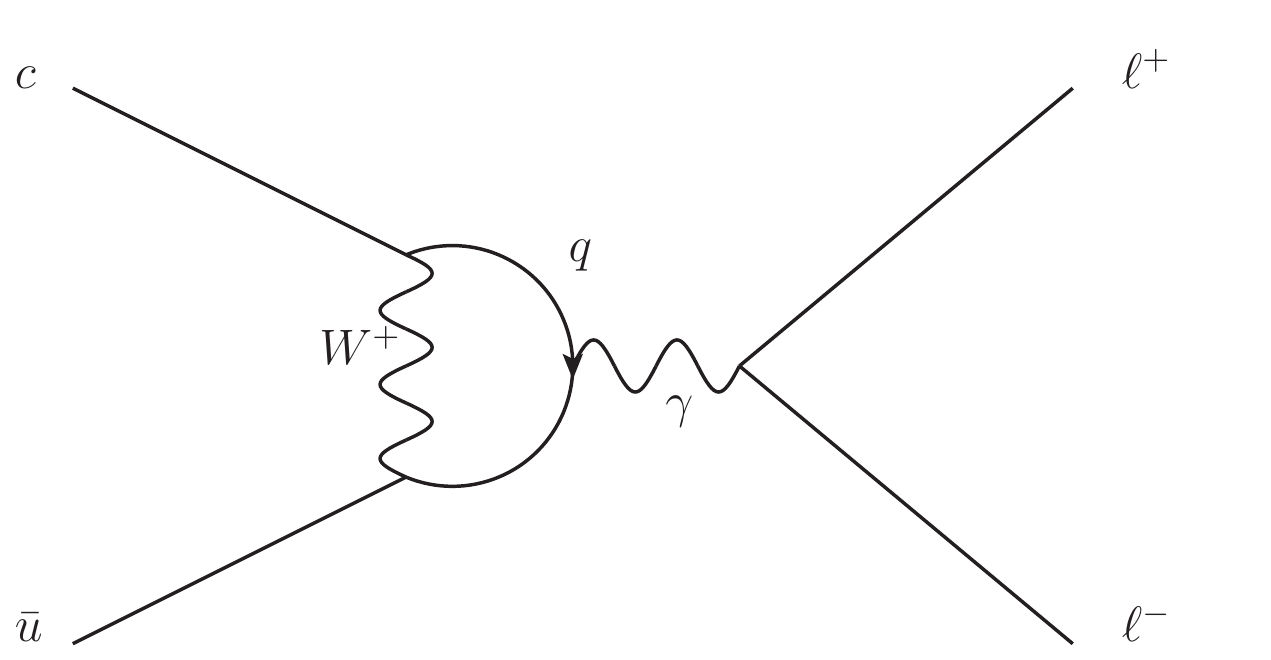}
    \caption{\label{fig:fcnc_sd}Loop level feynman diagrams for $D^0\to\ell^+\ell^-$.}
\end{figure}

Among all of the rare charm leptonic and semileptonic decays, perhaps decays with FCNC are of the most concern since the loop diagrams (e.g.~, Fig.~\ref{fig:fcnc_sd} for $D^0\to\ell^+\ell^-$) are expected to receive large contribution from physics beyond the SM.
The BF of $D^0\to\mu^+\mu^-$ are expected to be enhanced by SUSY or leptoquark to $10^{-8}$~\cite{Golowich2009} and $10^{-7}$~\cite{Dorsner2009}.
Currently the strongest limit is set by LHCb to be~\cite{lhcb_d0tomumu}
$$\mathcal{B}(D^0\to\mu^+\mu^-)<6.2\times10^{-9} @90\%~{\rm C.~L.},$$
which can heavily constrain new physics beyond the SM.

For charm semileptonic FCNC decays, the BFs may be enhanced by long distance contribution mediated via vector mesons to $10^{-6}$~\cite{Paul2011,Fajfer2007,Cappiello2013}.
In 2016, LHCb reported the BF~\cite{lhcb_d0tokpimumu} 
$$\mathcal{B}(D^0\to K^-\pi^+\mu^+\mu^-)=(4.17\pm0.12\pm0.40)\times10^{-6},$$
which is consistent with this prediction.
Recently, BaBar reported the result for $D^0\to K^-\pi^+e^+e^-$~\cite{babar_d0tokpiee} and found that
$$\mathcal{B}(D^0\to K^-\pi^+e^+e^-)=(4.0\pm0.5\pm0.2\pm0.1)\times10^{-6}$$
in $\rho/\omega$ resonance region, which agrees with LHCb's measurement, and
$$\mathcal{B}(D^0\to K^-\pi^+e^+e^-)<3.1\times10^{-6} @90\%~{\rm C.~L.}$$
at the continuum region.

To search for new physics effect in semilptonic FCNC decays, one has to avoid the resonance region where the long distance contributions dominate.
Such analyses have been performed at LHCb for $D^0\to\pi^+\pi^-\mu^+\mu^-$, $D^0\to K^+K^-\mu^+\mu^-$~\cite{lhcb_d0tokkmumu}, and $\Lambda_c^+\to p\mu^+\mu^-$~\cite{lhcb_lambdactopmumu}.
While significant signals are observed in the resonance region, no evidence is observed in the continuum region, which is consistent with the SM prediction.

Another method to search for new physics is to look at the asymmetry effect, such as the forward-backward asymmetry, triple-product asymmetry and CP asymmetry.
These observables are expected to be symmetric in the SM while several percent of asymmetry may be expected in physics beyond the SM.
LHCb has searched for this kind of asymmetry in $D^0\to \pi^+\pi^-\mu^+\mu^-$ and $D^0\to K^+K^-\mu^+\mu^-$~\cite{lhcb_asym} and no asymmetry is observed at current statistics.

\section{Summary}
In summary, BESIII has improved the precision of decay constants, form factors and CKM matrix elements in the charm sector with recent measurements.
Meanwhile, LFU test at a very high precision (1.5\% for Cabbibo favoured decays and 4\% for Cabbibo suppressed decays) has been performed while no evidence of violation is found.
Search for charm semileptonic decays to scalar mesons were performed at BESIII and the current results are in favor of the SU(3) nonet tetraquark description of $a_0(980)$, $f_0(500)$ and $f_0(980)$.
Moreover, our sensitivity to rare charm leptonic and semileptonic decays has been improved by several magnitudes with the huge statistics at LHCb, and strong constraints have been set for various new physics models with recent measurements.
With more data coming from BESIII, LHCb and BelleII, experiment study of charm leptonic and semileptonic decays will be further improved in the future.

\begin{acknowledgments}
The author thanks for the support by Joint Large-Scale Scientific Facility Funds of the National Natural Science Foundation of China and the Chinese Academy of Sciences under Contract No. U1532257.
\end{acknowledgments}

\bigskip 

\end{document}